\documentclass{iau} 
\usepackage{graphicx}

\title[IAUS 318.~~H\&R collision debris] 
{Erosive Hit-and-Run Impact Events: \\ Debris Unbound}

\author[Sarid, Stewart \& Leinhardt]   
{Gal Sarid$^1$,
Sarah T. Stewart$^2$
 \and Zo\"{e} M. Leinhardt$^3$}

\affiliation{$^1$Florida Space Institute, University of Central Florida, \\ 
Partnership 1 Building, 12354 Research Parkway, Suite 214, Orlando, FL, USA 32826-0650 \\ 
email: {\tt Gal.Sarid@ucf.edu} \\[\affilskip]
$^2$Department of Earth and Planetary Sciences, University of California Davis, \\
One Shields Avenue, Davis, CA USA 95618 \\ 
email: {\tt sts@ucdavis.edu} \\[\affilskip]
$^3$School of Physics, University of Bristol, \\
H.H. Wills Physics Laboratory, Tyndall Avenue, Bristol, BS8 1TL, UK \\
email: {\tt Zoe.Leinhardt@bristol.ac.uk}}

\pubyear{2015}
\volume{318}  
\jname{Asteroids: New Observations, New Models}
\editors{S. Chesley, R, Jedicke, A. Morbidelli \& D. Farnocchia, eds.}
\begin{document}

\maketitle

\begin{abstract}
Erosive collisions among planetary embryos in the inner solar system can lead to multiple remnant bodies, varied in mass, composition and residual velocity.  Some of the smaller, unbound debris may become available to seed the main asteroid belt. The makeup of these collisionally produced bodies is different from the canonical chondritic composition, in terms of rock/iron ratio and may contain further shock-processed material. Having some of the material in the asteroid belt owe its origin from collisions of larger planetary bodies may help in explaining some of the diversity and oddities in composition of different asteroid groups.      

\keywords{asteroids, minor planets, solar system: formation, methods: numerical}
\end{abstract}


\firstsection 
\section{Introduction}

The late stages of planet formation in regions closer to the Sun were characterized by vigorous dynamical interactions (e.g., \cite[Agnor et al. 1999]{Agnor_etal1999}  \cite[Raymond et al. 2009]{Raymond_etal2009}, \cite[Walsh et al. 2011]{Walsh_etal2011}, \cite[Chambers 2013]{Chambers2013}). This environment was conducive to giant impacts among planetary embryos. During this epoch bodies collide to eventually form the terrestrial planets, with impact velocities that could reach several times the escape velocity, in the range of several 10's of km/s.  

Outcomes of giant impacts include merging, grazing, partial accretion or erosion and catastrophic disruption, and depend on the specific dynamical conditions for each collision event (\cite[Asphaug 2010]{Asphaug2010}, \cite[Leinhardt \& Stewart 2012]{LeinhardtStewart2012}). Among these collision events the greatest potential for producing diverse end-members lies in the erosive Hit-and-Run (H\&R) regime (small mass ratios, off-axis oblique impacts and non-negligible ejected mass). This regime is also more probable in terms of the dynamical conditions for two planetary embryos colliding in the inner solar system (\cite[Asphaug 2010]{Asphaug2010}). 

An erosive collision regime has been invoked to explain several outstanding issues in the inner solar system, such as the formation of the Moon (\cite[Canup 2012]{Canup2012}, \cite[Cuk \& Stewart 2012]{CukStewart2012}), Mars' crustal dichotomy (e.g., \cite[Marinova et al. 2008]{Marinova_etal2008}), Mercury's origin by mantle stripping (e.g., \cite[Benz et al. 2007]{Benz_etal2007}, \cite[Sarid et al. 2014]{Sarid_etal2014}), planetary atmosphere and volatile loss (e.g., \cite[Genda \& Abe 2003]{GendaAbe2003}, \cite[Stewart et al. 2014]{Stewart_etal2014}). 

Energetic collision events have also been invoked to explain some of the diverse features seen in the main belt and meteoritic record. Among these are the presumed parent bodies of iron meteorites (both IVA iron and unclassified), which correspond to 50-100 differentiated and disrupted asteroid cores (\cite[Burbine et al. 2002]{Burbine_etal2002} , \cite[Goldstein et al. 2009]{Goldstein_etal2009}). The main-group pallasites represent another oddity, since these seem to have formed inside a mixed olivine and metallic iron-nickel body that formed due to interaction of a separated core and mantle material after a grazing collision (\cite[Yang et al.  2010 ]{Yang_etal2010}, \cite[Tarduno et al. 2012]{Tarduno_etal2012}). 

16 Psyche, being the most massive M-type asteroid, is suspected to be the completely stripped iron core of a larger differentiated parent body, but with no associated family to represent the ejected mantle material (\cite[Davis et al. 1999]{Davis_etal1999}). The presence of too few iron cores in the main belt, together with the lack of observational and meteoritic record for an expected large population of olivine-dominated asteroids (\cite[DeMeo \& Carry 2013]{DemeoCarry2013}), may indicate early collision scenarios that were set outside the current main belt. Indeed, dynamical pathways exist for remnant material from the inner solar system (even inside of 1 AU) outward to be implanted as main belt asteroids (e.g., \cite[Bottke et al. 2006]{Bottke_etal2006} or the ``Grand Tack'' scenario by \cite[Walsh et al. 2011]{Walsh_etal2011}).      

The work reported here is an analysis of part of a large ``giant impact'' simulation set, which is aimed at investigating the physical characteristics of remnant bodies after planetary embryo collision events. It is part of \cite[Sarid et al. (2014)]{Sarid_etal2014} and the supporting work is detailed in \cite [Sarid et al. (In Preparation)]{Sarid_etalPrep}. We note that while work has been done on analyzing the scaling of the largest remnant (e.g., \cite[Leinhardt \& Stewart 2012]{LeinhardtStewart2012}), the smaller remnants and debris have received less attention. 

When fragmentation is considered in full 3D numerical collision models, the produced debris clouds are and multiple small remnants are diverse (e.g., \cite[Asphaug 2010]{Asphaug2010}). We focus here only on the fate of the debris cloud of ejected material from a collision event, in terms of released mass and ejection velocity. These objects can be considered as proxies for some of the parent bodies in the early epoch of the main belt's evolution. Understanding the ``odd-balls'', some of which are targets for on-going and up-coming space missions and observations, is important because it gives us constraints about the formation and early evolution mechanisms in the inner solar system. 


\section{Methodology}

We performed and analyzed a set of simulations for collision events, covering a range in projectile and target mass ratios of 1 to 0.2 and impact velocities of about 2 to 5 times the mutual escape velocity (defined with the pre-impact target and projectile bodies). The impact angle has an important effect on the collision outcome. It is defined as the angle between the line connecting the centers of the two colliding bodies and the normal to the projectile velocity vector (see \cite[Leinhardt \& Stewart 2012]{LeinhardtStewart2012}). We varied this parameter between $0^{\circ}$ and $75^{\circ}$, corresponding to events of head-on collisions to barely grazing encounters.

We used the smoothed particle hydrodynamics (SPH) code \textit{GADGET 2} (\cite[Springel 2005]{Springel2005}), which was modified to handle tabulated equations of state, such as ANEOS (parameters used for SiO2, as mantle rock, and Fe, as core iron). This version of the SPH code has been previously used to successfully simulate planetary giant impacts (\cite[Marcus et al. 2009]{Marcus_etal2009}, \cite[Leinhardt et al. 2010]{Leinhardt_etal2010}, \cite[Cuk \& Stewart 2012]{CukStewart2012}).

The initial masses of bodies were set between 0.1 and 0.6 Earth mass ($M_{\oplus}$). Initial composition was set to 70\% silicate and 30\% iron, to reflect a roughly chondritic abundance starting point. Prior to impact runs, all bodies were allowed to initially settle to negligible particle velocities in isolation, within $\sim20$ simulated hrs, so that a planetary structure of core and mantle could be verified (\cite[Marcus et al. 2009]{Marcus_etal2009}). The total number of particles involved in each of our collision simulations was between $10^5$ and $3\times10^5$. All runs were followed for over 24 hrs of simulated post-impact time, in order for us to track the collisional shock processing and the provenance of material components and collision debris. Resulting configurations include stripped mantles, melting or vaporization of rock mantles and some iron core fractions,  and heavily disrupted and varied smaller remnants. The latter may be proxies for asteroid parent bodies, which acquire non-chondritic compositions even before going through collisional evolution in the main belt.

The choice of initial conditions for our collision scenarios covers much of the leading parameter ranges found in modern N-body simulations of terrestrial planet formation. Planetary embryo masses can initially range between 0.005 to 0.1 $M_{\oplus}$ and evolve to span most of the mass range of 0.1 to 0.6 $M_{\oplus}$, within the first $\sim100$ Myr of dynamical evolution (e.g, \cite[Raymond et al. 2009]{Raymond_etal2009}, \cite[Walsh et al. 2011]{Walsh_etal2011}). We note that not only are the masses consistent, but also the above mentioned impact velocities ($>$ twice the mutual escape velocity) are met, although higher velocities are less common throughout the different N-body simulation implementations (see \cite[Stewart \& Leinhardt 2012]{StewartLeinhardt2012}). Thus, our chosen mass spectrum is not special and the velocity spectrum is quite plausible. 


We note here that because of particle resolution in our SPH simulations the smallest gravitationally bound remnant body has a mass $\gtrsim1.5 M_{Ceres}$. The debris cloud represents the sub-resolution mass, which has potential energy greater than the smallest remnant body. The latter is any gravitational aggregate greater than 100 SPH particles or more, in order to avoid smoothing length biases. 

At the end of each simulation we record the total mass of the debris cloud and the mass of each material type, which is either rock or iron (following the initial designation of equation of state, for each SPH particle). By doing so, we can calculate $v_{RMS}$ as the mean velocity of the debris cloud. This mass-weighted velocity is relative to the center of mass of the collision event.


\section{Results}

\begin{figure}[htbp]
\centering
\begin{minipage}{0.5\textwidth}
  \centering
  \includegraphics[width=1.0\linewidth]{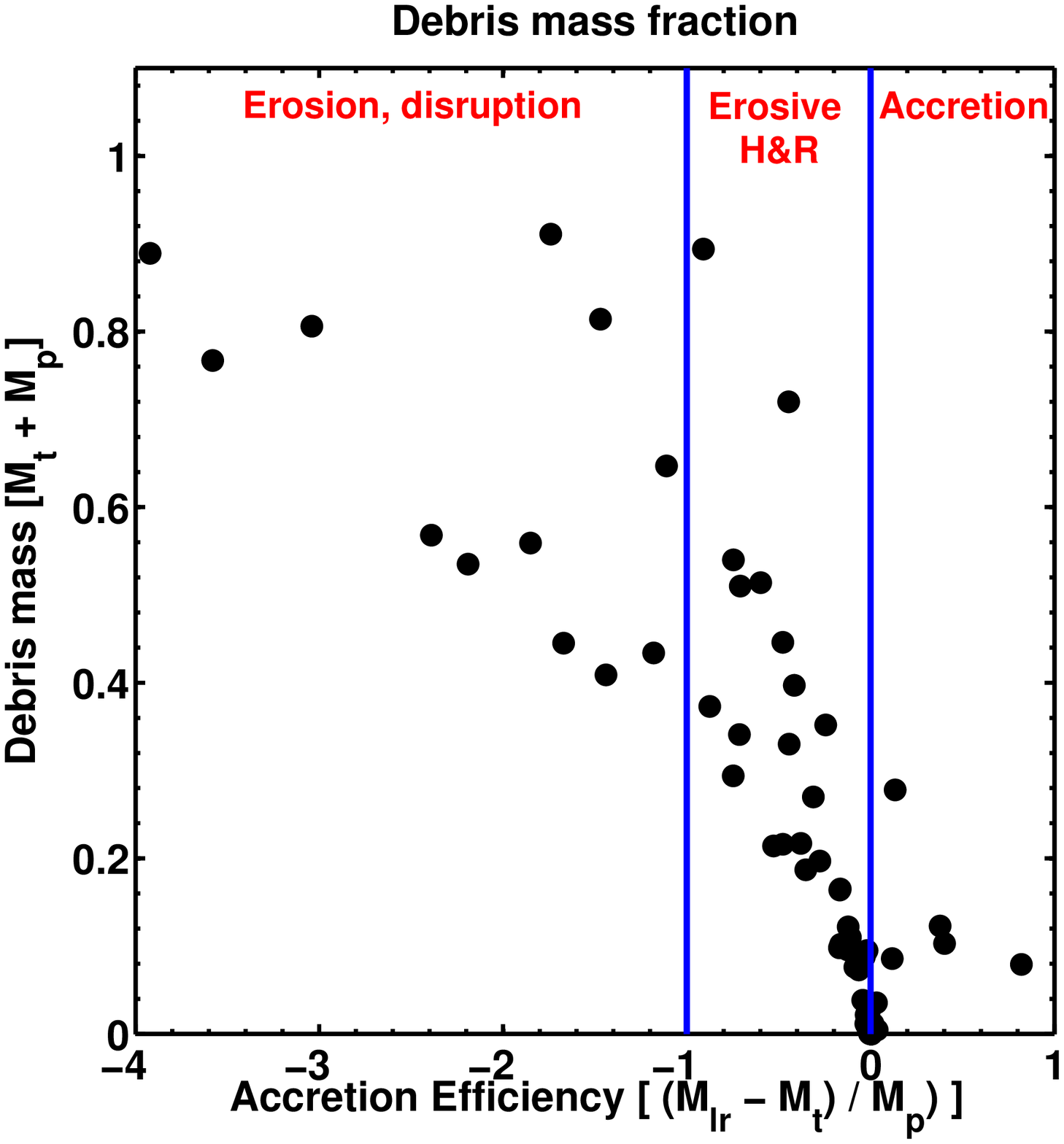}
  \parbox{0.92\linewidth}{\caption{Debris mass distribution as a function of the accretion efficiency parameter. Mass is shown as fraction of total initial mass (target and projectile) and vertical lines separate the collision regimes. Each data point is an impact simulation outcome of total debris mass (silicate-mantle and iron-core). \label{MassDist}}}
\end{minipage}%
\begin{minipage}{0.5\textwidth}
  \centering
  \includegraphics[width=1.0\linewidth]{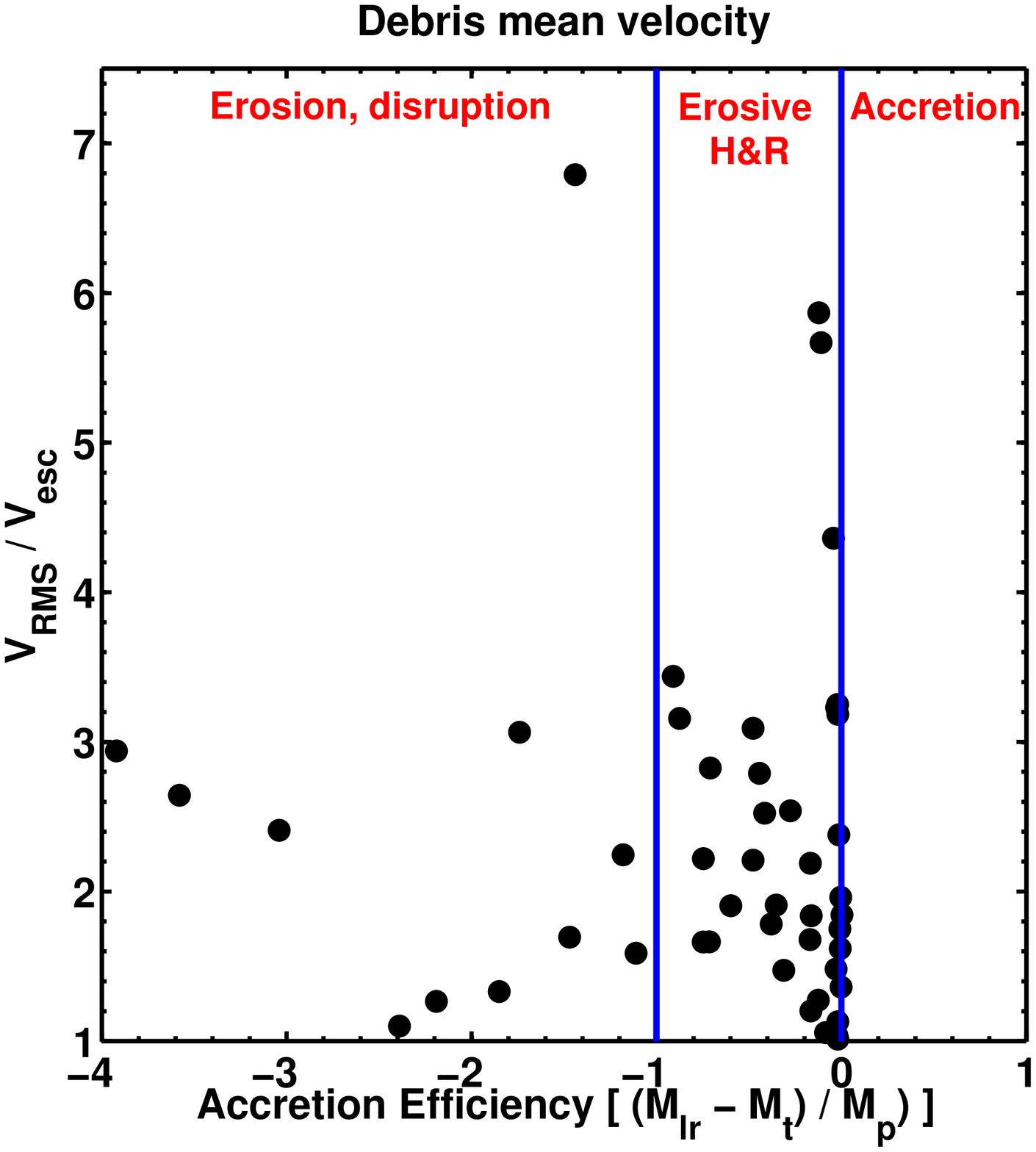}
  \parbox{0.92\linewidth}{\caption{RMS velocity distribution as a function of the accretion efficiency parameter. Velocities are normalized to the largest remnant's escape velocity and vertical lines separate the collision regimes. Each data point is an impact simulation outcome for which the debris clouds are escaping re-accretion. \label{VrmsDist}}}
\end{minipage}%
\end{figure}

Fig.\,\ref{MassDist} shows the mass of each debris cloud as a fraction of the total initial mass of the system (target and projectile). The total mass of the debris is shown as a function of the accretion efficiency, defined as $(M_{lr}-M_t)/M_p$ (\cite[Asphaug 2010]{Asphaug2010}). The mass subscripts are for the largest remnant (``lr''), the original target (``t'') and the original projectile (``p''). This representation clearly identifies the different regimes and is easy to determine from collision calculations (\cite[Asphaug 2010]{Asphaug2010}, \cite[Leinhardt \& Stewart 2012]{LeinhardtStewart2012}). Each point represents the end result of a full simulation of a specific impact scenario (mass ratio, impact velocity and impact angle). 

The erosive H\&R regime, between the two vertical lines, covers a transition from net loss (erosion) to net gain (accretion) of material by the target. Thus, it has a more complex interaction of the target and projectile bodies (their mantles and cores) with the near-field and far-field gravitational potential. It exhibits a larger scatter in the mass released into the debris cloud and varying silicate-to-iron ratios. This ratio, which represents the contribution of mantle and core material to the debris cloud respectively, varies between an all mantle composition to roughly 75\% silicate. The larger fraction of iron in the debris correlates with the most energetic collisions, involving less grazing impact angles ($\lesssim35^\circ$) and higher impact velocities ($\gtrsim30$ km/s).  

The total mass in the debris clouds may exceed that of the asteroid main belt and is temporarily unbound. However, not all of the released mass will actually remain as debris cloud for any extended time. If the RMS velocity is smaller than the escape velocity of the largest remnant in the post-impact system, then these debris will be re-accreted within a few orbital timescales at most (\cite[Asphaug 2010]{Asphaug2010}). 

Fig.\,\ref{VrmsDist} shows the distribution of RMS velocities (the mean velocities of the debris clouds), normalized to the escape velocity of the largest remnant, $v_{esc}$. This object is the most massive in the post-impact system and has the largest potential to sweep-up debris. Since the largest remnant is also an outcome of the collision event, its escape velocity is calculated from each simulation and ranges between about 2.5 to 10 km/s. The smaller escape velocities correspond to the disruptive regime (smaller remnant masses). We only show cases where the velocity ratio is greater than 1, because all the other cases will not have the potential to achieve any dynamical separation from the collision remnants. 

The erosive H\&R regime contains the largest scatter and high RMS values, relative to the largest remnant's escape velocity. Thus, impact events in this regime are more likely to produce not only a large scatter in mass and related composition, but also scatter in the debris cloud velocity field. Subsequently, debris material may have a higher probability of surviving on orbital timescales, before being re-accreted onto the largest remnant. 

However, not all of these escaping debris will survive being eventually re-accreted, as their trajectories may become orbit-crossing with the post-impact remnants (target or projectile). In order to assess whether the debris clouds (rock or iron) can eventually escape the collision system's center of mass, we employ another parametrization of the velocity field. 

We introduce a notion of critical velocity to escape the feeding zone, following and modifying the analysis of \cite[Margot (2015)]{Margot2015}. This approximates the criterion for the central body (the largest remnant of the post-impact system) to clear all debris from its orbital feeding zone, based on energy considerations and an idealized instantaneous velocity kick to the debris. 

The energy that debris require in order to clear the feeding zone is given by: $\Delta\epsilon = (\mu/2)/a_S - (\mu/2)/(a_S+CR_{H,S})$. Here $\mu$ is the gravitational parameter $GM$, where $G$ is the gravitational constant and $M = M_{\odot}+M_{lr}$, with $M_{lr}$ as the mass of the largest remnant (the post-impact target in non-disruptive collisions). The heliocentric distance for the collision system's center of mass is $a_S$. The effective Hill radius is $R_{H,S} = a_S(M_{lr}/3M_{\odot})^{1/3}$. The numerical constant $C$ is usually taken as $2\sqrt{3}$, when considering clearing of a planet feeding zone (e.g., \cite[Lissauer 1993]{Lissauer1993}). 

We can now rearrange terms and consider that even for a perfect merging scenario, where the largest remnant is just the sum of target and projectile masses, it holds that $M_{lr} \ll M_{\odot}$. The resulting expression for the orbital clearing energy is: $\Delta\epsilon = (C\mu/2)R_{H,S}/a_S^2$. We can compare this with the orbital energy of an object moving at an orbit $a_S$ and having a velocity $v_{kick}$: $\epsilon_{orb} = v_{kick}^2/2 - \mu/a_S$.  

By comparing $\Delta\epsilon = \epsilon_{orb}$, we can define a critical velocity $v_{kick}^*$ that a debris cloud must achieve in order to separate from the post-impact remnants on a shorter timescale than the orbital timescale. This derived threshold value is not relative to the center of mass of the collision system, so we have to correct it by subtracting the keplerian velocity $v_S = \sqrt{GM_{\odot}/a_S}$. 

The debris cloud's critical velocity to clear the feeding zone of the post-impact system is given by: $v^* = v_{kick}^* - v_S = v_{\oplus}\sqrt{(1AU/a_S)}(\sqrt{f}-1)$, where $v_{\oplus}$ is Earth's orbital velocity. Here $f = [C(M_{lr}/3M_{\odot})^{1/3} + 2]$, but the first term can be neglected for all considered values of masses and reasonable values for the numerical constant. Thus, for an erosive H\&R impact at 1 AU, the debris cloud material has to be ejected from the target or projectile with an instantaneous RMS velocity of roughly 12.7 km/s. At this initial kick, it will be cleared from the feeding zone of the post-impact remnants and in effect be dynamically separated from the collision system.  

\begin{figure}[htbp]
\centering
  \centering
  \includegraphics[width=0.6\linewidth]{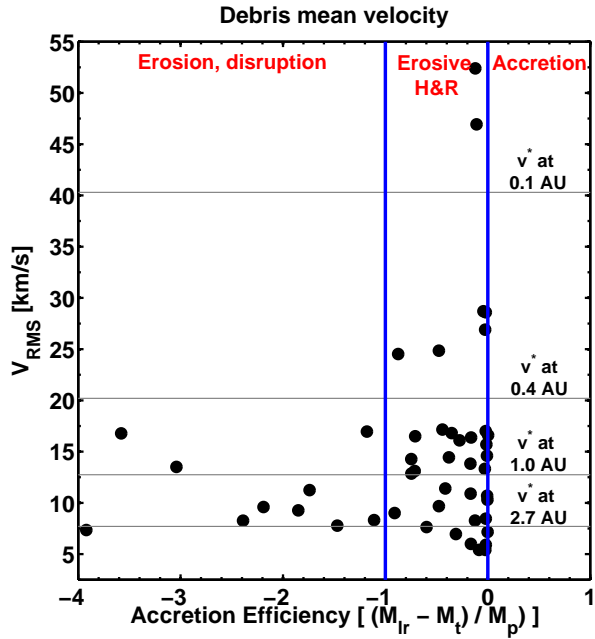}
  \parbox{0.92\linewidth}{\caption{RMS velocity distribution (rock-mantle in blue, iron-core in black), for each set of impact conditions, by collision regime (accretion efficiency parameter). Threshold values (horizontal lines) represent the critical velocity to escape a feeding zone centered at the designated heliocentric distance. \label{ThresholdVel}}}
\end{figure}

Fig.\,\ref{ThresholdVel} compares the distribution of RMS velocities with the critical orbit clearing velocity. The velocities are from the same subset of simulations as in Fig.\,\ref{VrmsDist}. The horizontal lines denote different critical values of $v^*$, corresponding to different heliocentric locations for the collision center of mass. 

For the planetary orbits of the inner solar system, we calculate that any debris cloud with a mean velocity $v_{RMS} \lesssim 10$ km/s (the critical velocity value at the orbit of Mars) will probably not be able to clear the orbital feeding zone of the largest remnant (the ``target''). If giant impacts occur at the main belt's mean orbit ($\sim2.7$ AU), the debris could escape the feeding zone of the post-impact remnant. For a nominal compact planetary system, where giant impacts can occur closer in (case of 0.1 AU), very few impact conditions can result in debris escaping from the feeding zone.  

If erosive H\&R events occurred at or inside of the orbit of Venus ($v^*\simeq15$ km/s), we expect a comparatively small contribution of dispersed and implanted debris into the main belt. Since most of the erosive H\&R impacts produce dominantly rock (mantle) debris, the contribution of core-dominated objects would have been even smaller. As a result, it would be scarcely represented in the observational and meteoritic records. This is without invoking any special mechanisms for the presumed paucity of metallic cores and lack of olivine asteroids, other than the the giant impact epoch already common for terrestrial planet embryos. 

We note that out of our set of simulations roughly 40\% resulted in multiple remnant bodies. This means that while the remnant target body has remained mostly unscathed, the projectile body has been eroded or disrupted altogether, with some of the ejected mass re-accreting quickly to form several smaller remnant bodies. 

Most of the smaller remnant bodies are made of silicate mantle material. These are products of either the projectile's blasted mantle layers or a mix of target and projectile mantles. The corresponding impact conditions include more massive targets and higher impact velocities for oblique impacts with impact angles between $40^{\circ}$ and $60^{\circ}$. 

However, a small number of iron core proxies ($> 95\%$ iron in the bulk composition), reminiscent of 16 Psyche or some of the M-type asteroids, can be produced as well. These bodies are derived only from the projectile iron core and are not directly ejected from the post-impact feeding zone. The corresponding collisions include the most massive targets and impact angles $< 40^{\circ}$. Thus, metallic asteroids might owe their origin to the most erosive H\&R collisions or even nearly head-on catastrophic disruptions of differentiated projectiles (smaller body in the collision system).



\section{Summary}
In the context of large planetary formation simulations, velocity and impact angle distributions are necessary to asses impact probabilities. The mass distribution and interaction within planetary embryo and asteroid swarms depends both on gravitational dynamics and the applied fragmentation mechanism.

We have discussed results from multiple collision simulations, which cover a wide range of target and projectile mass ratios, impact velocities and impact angles. Among these events, the erosive hit-and-run (H\&R) impacts are quite common in the context of current dynamical evolution scenarios in the inner solar system. The post-impact configuration is usually a pair comprising of a roughly unscathed ``target'' (larger remnant mass) and a heavily eroded secondary remnant mass, together with a cloud of debris. The latter could potentially escape from the post-impact large remnants and be implanted in other orbits. In some cases, we have also noted survival of small remnant bodies, which have a mixed composition of initial material and provenance. Thus, these erosive collision events can produce a diverse population of bodies smaller than terrestrial planetary embryos with non-chondritic compositions. The collision debris may become available early-on in solar system evolution to seed the main belt.
 
Higher-resolution simulations are being constructed and analyzed, in order to gain a better understanding of the size-dependence of remnant and debris compositions. Current on-going and future work focus on constructing a ``universal law'' (e.g., \cite[Leinhardt \& Stewart 2012]{LeinhardtStewart2012}, \cite[Stewart \& Leinhardt 2012]{StewartLeinhardt2012}), which will deal with general projectile remnant scaling relations (mass, velocity, composition) and the mass and velocity distribution of unbound debris. 




\begin{thebibliography}{}

\bibitem[Agnor et al. (1999)]{Agnor_etal1999} 
{Agnor, C.~B., Canup,  R.~M., \& Levison, H.~F.} 1999, 
\textit{Icarus}, 142, 219 

\bibitem[Asphaug (2010)]{Asphaug2010} 
{Asphaug, E.} 2010, 
\textit{Chemie der Erde / Geochemistry}, 70, 199 

\bibitem[Benz et al. (2007)]{Benz_etal2007} 
{Benz, W., Anic, A., Horner, J., \& Whitby, J.~A.} 2007, 
\textit{Sp. Sci. Rev.}, 132, 189 

\bibitem[Bottke et al. (2006)]{Bottke_etal2006} 
{Bottke, W.~F., Nesvorn{\'y}, D., Grimm, R.~E., Morbidelli, A., \& O'Brien, D.~P.} 2006, 
\textit{Nature}, 439, 821 

\bibitem[Burbine et al. (2002)]{Burbine_etal2002} 
{Burbine, T.~H., McCoy, T.~J., Meibom, A., Gladman, B., \& Keil, K.} 2002, 
\textit{Asteroids III}, (University of Arizona Press, Tucson), p.\, 653 

\bibitem[Canup (2012)]{Canup2012} 
{Canup, R.~M.} 2012, 
\textit{Science}, 338, 1052 

\bibitem[Chambers (2013)]{Chambers2013} 
{Chambers, J.~E.} 2013, 
\textit{Icarus}, 224, 43 

\bibitem[Consolmagno et al. (2015)]{Consolmagno_etal2015} 
{Consolmagno, G.~J., Golabek, G.~J., Turrini, D., et al.} 2015, 
\textit{Icarus}, 254, 190 

\bibitem[Cuk \& Stewart (2012)]{CukStewart2012} 
{Cuk, M., \& Stewart, S.~T.} 2012, 
\textit{Science}, 338, 1047 

\bibitem[Davis et al. (1985)]{Davis_etal1985} 
{Davis, D.~R., Chapman, C.~R., Weidenschilling, S.~J., \& Greenberg, R.} 1985, 
\textit{icarus}, 62, 30

 \bibitem[Davis et al. (1999)]{Davis_etal1999} 
{Davis, D.~R., Farinella, P., \& Marzari, F.} 1999, 
\textit{Icarus}, 137, 140  

\bibitem[DeMeo \& Carry (2013)]{DemeoCarry2013} 
{DeMeo, F.~E., \& Carry, B.} 2013, 
\textit{Icarus}, 226, 723 


\bibitem[Genda \& Abe (2003)]{GendaAbe2003} 
{Genda, H., \& Abe, Y.} 2003, 
\textit{Icarus}, 164, 149 

\bibitem[Goldstein et al. (2009)]{Goldstein_etal2009} 
{Goldstein, J.~I., Scott, E.~R.~D., \& Chabot, N.~L.} 2009, 
\textit {Chemie der Erde / Geochemistry}, 69, 293 

\bibitem[Leinhardt et al. (2010)]{Leinhardt_etal2010} 
{Leinhardt, Z.~M., Marcus, R.~A., \& Stewart, S.~T.} 2010, 
\textit{ApJ}, 714, 1789 

\bibitem[Leinhardt \& Stewart (2012)]{LeinhardtStewart2012} 
{Leinhardt, Z.~M., \& Stewart, S.~T.} 2012, 
\textit{ApJ}, 745, 79

\bibitem[Lissauer (1993)]{Lissauer1993} 
{Lissauer, J.~J.} 1993, 
\textit{ARA\&A}, 31, 129 

\bibitem[Marcus et al. (2009)]{Marcus_etal2009} 
{Marcus, R.~A., Stewart, S.~T., Sasselov, D., \& Hernquist, L.} 2009, 
\textit{ApJL}, 700, L118

\bibitem[Margot (2015)]{Margot2015} 
{Margot, J.-L.} 2015, 
arXiv:1507.06300 
  
\bibitem[Marinova et al. (2008)]{Marinova_etal2008} 
{Marinova, M.~M., Aharonson, O., \& Asphaug, E.} 2008, 
\textit{Nature}, 453, 1216 

\bibitem[Raymond et al. (2009)]{Raymond_etal2009} 
{Raymond, S.~N., O'Brien, D.~P., Morbidelli, A., \& Kaib, N.~A.} 2009, 
\textit{Icarus}, 203, 644 

\bibitem[Sarid et al. (2014)]{Sarid_etal2014} 
{Sarid, G., Stewart, S.~T., \& Leinhardt, Z.~M.} 2014, 
\textit{LPI Cont.}, 45, 2723

\bibitem[Sarid et al. (In Preparation)]{Sarid_etalPrep} 
{Sarid, G., Stewart, S.~T., \& Leinhardt, Z.~M.} In preparation

\bibitem[Springel (2005)]{Springel2005} 
{Springel, V.} 2005, 
\textit{MNRAS}, 364, 1105  

\bibitem[Stewart \& Leinhardt (2012)]{StewartLeinhardt2012} 
{Stewart, S.~T., \& Leinhardt, Z.~M.} 2012, 
\textit{ApJ}, 751, 32 

\bibitem[Stewart et al. (2014)]{Stewart_etal2014} 
{Stewart, S.~T., Lock, S.~J., \& Mukhopadhyay, S.} 2014, 
\textit{LPI Cont.}, 45, 2869 

\bibitem[Tarduno et al. (2012)]{Tarduno_etal2012} 
{Tarduno, J.~A., Cottrell, R.~D., Nimmo, F., et al.} 2012, 
\textit{Science}, 338, 939 

\bibitem[Walsh et al. (2011)]{Walsh_etal2011} 
{Walsh, K.~J., Morbidelli, A., Raymond, S.~N., O'Brien, D.~P., \& Mandell, A.~M.} 2011, 
\textit{Nature}, 475, 206 

\bibitem[Yang et al. (2010)]{Yang_etal2010} 
{Yang, J., Goldstein, J.~I., \& Scott, E.~R.~D.} 2010, 
\textit{Geochim. Cosmochim. Acta}., 74, 4471 

\end{thebibliography}
\end{document}